\documentclass[preprint,preprintnumbers,amsmath,amssymb,superscriptaddress]{revtex4}

\usepackage{graphicx}% Include figure files
\usepackage{dcolumn}% Align table columns on decimal point
\usepackage{bm}% bold math

\usepackage{graphicx}% Include figure files
\usepackage{dcolumn}% Align table columns on decimal point
\usepackage{amsmath}    % need for subequations
\usepackage{verbatim}   % useful for program listings
\usepackage{color}      % use if color is used in text
\usepackage{subfigure}  % use for side-by-side figures
\usepackage{hyperref}   % use for hypertext links, including those to external documents and URLs
\usepackage{amsfonts}
\usepackage{bm}% bold math

\begin{document}

%\preprint{Toshiba Research Europe Limited/Confidential}
\title[Sample title]{Cavity-enhanced coherent light scattering from a quantum dot}% Force line breaks with \\
\author{A. J. Bennett}
\affiliation{Toshiba Research Europe Limited, Cambridge Research Laboratory, 208 Science Park, Milton Road, Cambridge, CB4 0GZ, U. K.}%Lines break automatically or can be forced with \\
\email{anthony.bennett@crl.toshiba.co.uk}

\author{J. P. Lee}%
\affiliation{Toshiba Research Europe Limited, Cambridge Research Laboratory, 208 Science Park, Milton Road, Cambridge, CB4 0GZ, U. K.}%Lines break automatically or can be forced with \\
\affiliation{ Engineering Department, University of Cambridge, 9 J.J. Thomson Avenue, Cambridge, CB3 0FA, U.K. %\\This line break forced with \textbackslash\textbackslash
}%

 \author{D. J. P. Ellis}
\affiliation{Toshiba Research Europe Limited, Cambridge Research Laboratory, 208 Science Park, Milton Road, Cambridge, CB4 0GZ, U. K.}%Lines break automatically or can be forced with \\

 \author{T. Meany}
\affiliation{Toshiba Research Europe Limited, Cambridge Research Laboratory, 208 Science Park, Milton Road, Cambridge, CB4 0GZ, U. K.}%Lines break automatically or can be forced with \\

 \author{E. Murray}
\affiliation{Toshiba Research Europe Limited, Cambridge Research Laboratory, 208 Science Park, Milton Road, Cambridge, CB4 0GZ, U. K.}%Lines break automatically or can be forced with \\
\affiliation{Cavendish Laboratory, Cambridge University,\\
J. J. Thomson Avenue, Cambridge, CB3 0HE, U. K.}

 \author{F. Fl\"{o}ther}
\affiliation{Toshiba Research Europe Limited, Cambridge Research Laboratory, 208 Science Park, Milton Road, Cambridge, CB4 0GZ, U. K.}%Lines break automatically or can be forced with \\
\affiliation{Cavendish Laboratory, Cambridge University,\\
J. J. Thomson Avenue, Cambridge, CB3 0HE, U. K.}

\author{J. P. Griffiths}
\affiliation{Cavendish Laboratory, Cambridge University,\\
J. J. Thomson Avenue, Cambridge, CB3 0HE, U. K.}

\author{I. Farrer}
\affiliation{Cavendish Laboratory, Cambridge University,\\
J. J. Thomson Avenue, Cambridge, CB3 0HE, U. K.}

\author{D. A. Ritchie}
\affiliation{Cavendish Laboratory, Cambridge University,\\
J. J. Thomson Avenue, Cambridge, CB3 0HE, U. K.}

\author{A. J. Shields}
\affiliation{Toshiba Research Europe Limited, Cambridge Research Laboratory, 208 Science Park, Milton Road, Cambridge, CB4 0GZ, U. K.}%Lines break automatically or can be forced with \\

\date{\today}% It is always \today, today,

%\begin{abstract}
%there is no abstract for NPhoton
%\end{abstract}

\pacs{Valid PACS appear here}% PACS, the Physics and Astronomy
%Use showkeys class option if keyword

\begin{abstract}
Resonant excitation of atoms and ions in macroscopic cavities has
lead to exceptional control over quanta of light \cite{Haroche06,
Wineland13}. Translating these advantages into the solid state with
emitters in micro-cavities promises revolutionary quantum
technologies in information processing and metrology. Key is
resonant optical reading and writing from the emitter-cavity system.
However, it has been widely expected that the reflection of a
resonant laser from a micro-fabricated $\lambda^{3}$-sized cavity
would dominate any quantum signal. Here we demonstrate coherent
photon scattering from a quantum dot in a micro-pillar. The cavity
is shown to enhance the fraction of light which is resonant
scattered towards unity, generating anti-bunched indistinguishable
photons a factor of 16 beyond the time-bandwidth limit, even when
the transition is near saturation. Finally, deterministic excitation
is used to create 2-photon N00N states with which we make
super-resolving phase measurements in a photonic circuit.
\end{abstract}

\maketitle

Engineering cavities around an emitter modifies the local density of
optical states, changing the emission pattern and radiative decay
rate \cite{Buckley14}. It has been proposed that cavities can also
accelerate the rate at which a spin may be prepared \cite{Loo11},
increase photon-spin coupling \cite{Hu08} and enhance Raman
scattering \cite{Sweeney14} under resonant optical fields. A reduced
radiative lifetime $T_{1}$ also leads to an increase in the photon
fraction that can be resonantly scattered \cite{Nguyen11,
Konthasinghe12} leading to an ``ideal" quantum light source with
high efficiency and high coherence.

Three dimensional pillar micro-cavities allow a cavity-induced
reduction in $T_{1}$ by an order of magnitude \cite{Gerard98,
Gazzano13, Ulhaq12, Ates09, Santori02, Pooley12, Moreau01}. Unlike
previous studies \cite{Ates09, Ulhaq12} we excite and collect
photons from the cavity along the axis that couples efficiently to
the light field, illustrating the potential of this system as a
spin-photon interface and a source of indistinguishable single
photons.

\begin{figure}[h]
\includegraphics[width=80mm]{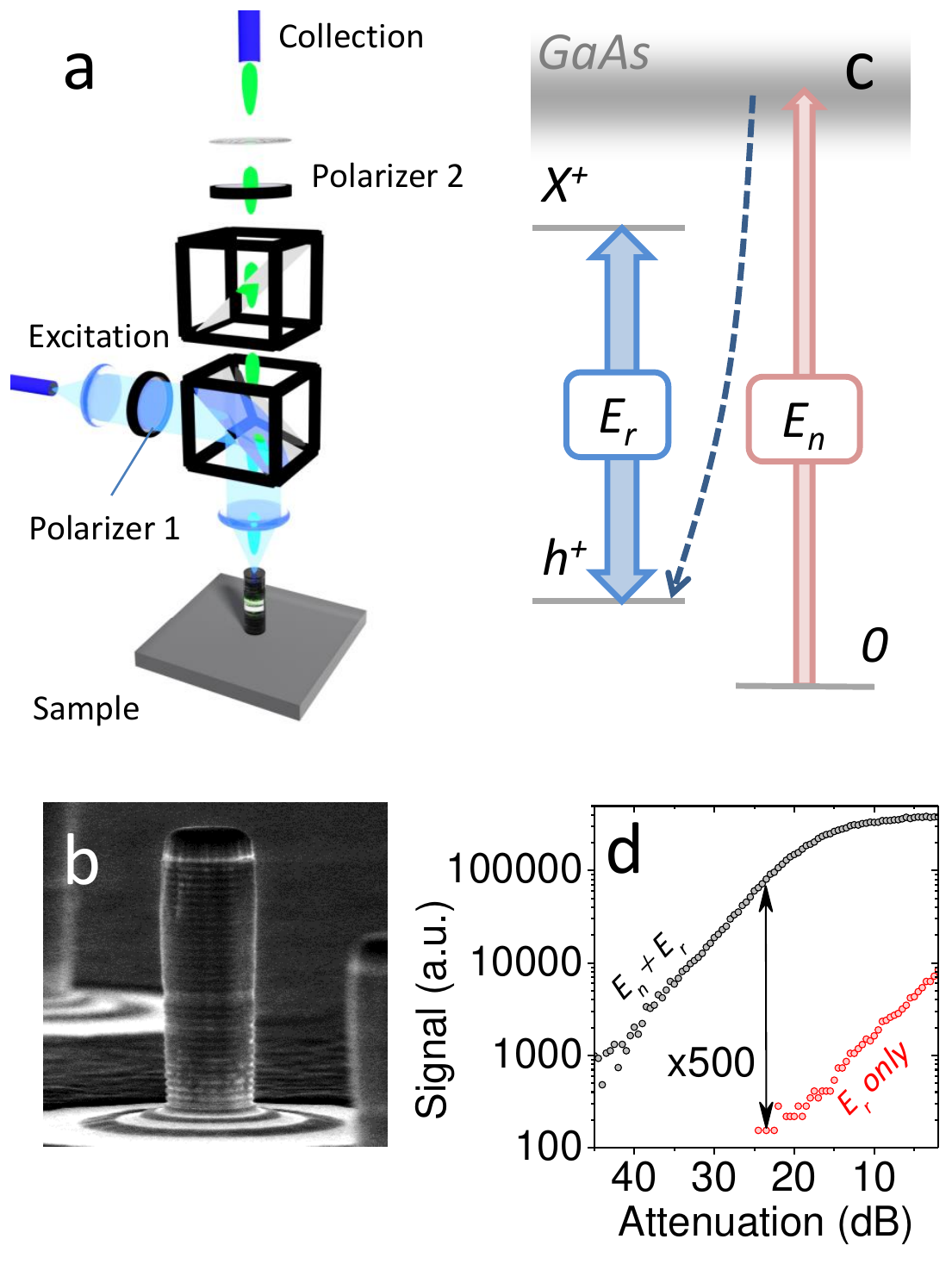}% Here is how to import EPS art
\caption{\label{Fig1} \textbf{Resonant excitation of a quantum dot
in a micro-cavity} (a) Experimental arrangement (b) SEM image of of
a pillar micro-cavity. (c) Energy level diagram with the resonant
driving field $E_{r}$ coupled to the $X^{+}$ transition. Resonant
emission can be gated with the weak non-resonant laser, $E_{n}$
which creates electron-hole pairs in the GaAs band-gap. Following
hole capture in the dot (dashed arrow) the $X^{+}$ transition may be
resonantly driven. (d) Power dependence of the source intensity as a
function of the coherent laser intensity, with and without
additional excitation by the weak non-resonant laser, $E_{n}$. }
\end{figure}

It is possible to suppress the laser signal at the detectors whilst
efficiently collecting the emission from an etched micro-cavity
using polarisation filtering, Figure \ref{Fig1}a. Thus far, this
technique has been limited to optically smooth and flat samples
\cite{Matthiesen12, Kuhlmann13, He13}, for which a cavity
enhancement is not observed. We employ a pillar micro-cavity (Figure
\ref{Fig1}b)
 with a 2.25 $\mu$m diameter, which is close to optimal for
maximizing the photon extraction efficiency \cite{Pelton02,
Gazzano13}. The $HE_{11}$ mode of the device we study has a quality
factor, $Q$, of 8,900 which is reduced from the $Q$ of the unetched
cavity. Imperfections in the cavity sidewalls which are visible in
Figure \ref{Fig1}b are a possible source of optical loss in the mode
\cite{Gerard98, Pelton02}. Rotating polariser 2 in the photon
detection path it was possible to suppress the laser collected by a
factor of 10$^{3}$.

Figure \ref{Fig1}d shows data from this micro-cavity, which contains
an $X^{+}$ transition emitting at 934 nm. Under only resonant
illumination (the laser denoted $E_{r}$ in Figure \ref{Fig1}c) we
collect a total absence of emission from $X^{+}$, only weak
poissonian laser (red data points, Figure \ref{Fig1}d). However, the
addition of $<$ 100 pW of illumination at 850 nm (laser $E_{n}$)
``activates" the resonant emission (RE). We attribute this to the
capture of a single hole into the ground state by the process shown
in Figure \ref{Fig1}c. This is to be contrasted with the optical
gating of neutral excitons, in which it has been reported RE can be
suppressed when a charge tunnels into the dot from a nearby defect
\cite{Nguyen12}. The ability to switch on the RE with this weak
non-resonant light enables us to quantify the intensity of the RE
relative to the laser. For the device discussed here the RE
collected is 500 times greater than the laser (Figure \ref{Fig1}d),
but can reach a factor of 3000 in other cavities.

The resonant laser power required to observe RE in these high-$Q$
cavities is $\sim$ 3 orders of magnitude lower than for planar
cavities with $Q \sim$70 \cite{Lee15}. This is to be expected given
the efficient photon-in-coupling and higher quality factor. In
addition, the Purcell effect has reduced the lifetime of the single
quantum emitter, broadening the transition to $\triangle E =$ 6.14
$\pm$ 0.19 $\mu$eV (approximately 5 times greater than non-cavity
enhanced emitters in this sample).

\begin{figure}[h]
\includegraphics[width=160mm]{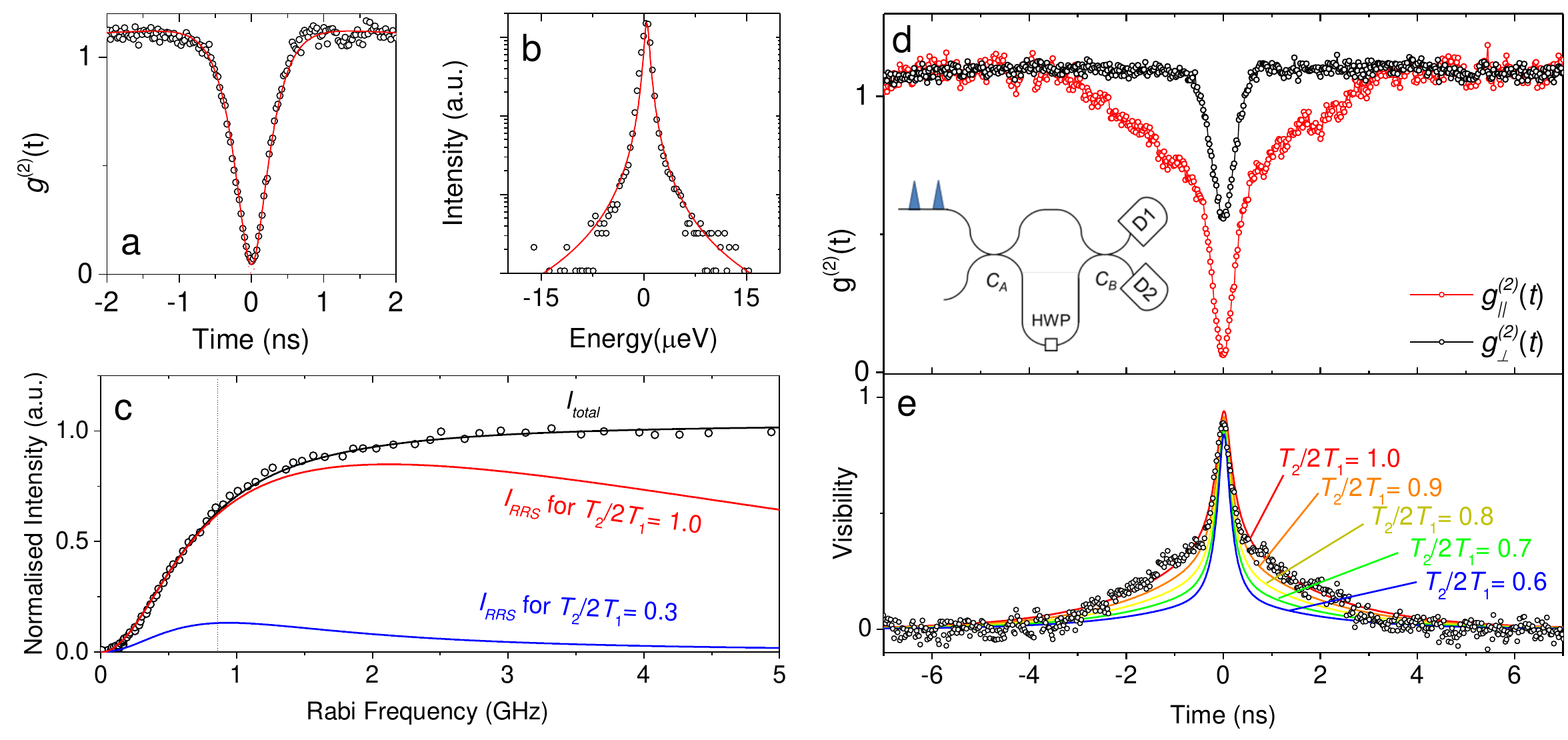}% Here is how to import EPS art
\caption{\label{Fig2} \textbf{Cavity-enhanced Resonant Rayleigh
Scattering} (a) Auto-correlation measurement at a Rabi frequency of
0.83 GHz and (b) spectrum of the emitted light at the same power
(black data points) with instrument resolution (red). (c) The power
dependence of the emission (black) is shown as a function of Rabi
frequency. From this the proportion of the light due to Resonant
Rayleigh scattering is calculated for $T_{2}/2T_{1}$ = 1.0 and 0.3.
(d) Post-selected Hong-Ou-Mandel auto-correlation for parallel (red)
and orthogonal photon polarisations (black). (e) Interference
visibility deduced from (d), fitted with different values of
$T_{2}/2T_{1}$.}
\end{figure}

Figure \ref{Fig2}a shows the result of a Hanbury-Brown and Twiss
auto-correlation measurement of $g^{(2)}(t)$ recorded under CW
excitation at a Rabi frequency $\Omega =$ 0.83 GHz. The data is
fitted with the well-known equations for $g^{(2)}(t)$ under coherent
excitation \cite{Cohen98, Nguyen12, Flagg09}, including an
additional charging-induced bunching effect \cite{Santori04}. This
confirms the dominance of the anti-bunched quantum emission at the
detectors. The spectrum of the RE in Figure \ref{Fig2}b (black data
points), appears close to the instrument resolution (0.78 $\mu$eV,
red line). There is no evidence of the emitter linewidth in this
spectrum. This is a clear signature of Resonant Rayleigh Scattering
(RRS) by the two level system. A least-squares fit gives a spectral
width of 0.37 $\pm$ 0.03 $\mu$eV over the system response, so the
coherently scattered photons are narrower than the radiative
linewidth by a factor of 16.

Figure \ref{Fig2}c shows the total intensity, $I_{total}$, emitted
by the transition as a function of Rabi frequency. One advantage of
cavity-enhancement is that emission due to RRS, $I_{RRS}$, is
greatly increased. The fraction of the total emitted light due to
RRS \cite{Cohen98, Nguyen11} is

\begin{equation}\label{equ1}
\frac{I_{RRS}}{I_{total}} =
\frac{T_{2}}{2T_{1}(1+\Omega^{2}T_{1}T_{2})}
\end{equation}
\\
where $T_{1}$ is the radiative lifetime and $T_{2}= 2\hbar/\triangle
E$. For an emitter with no cavity enhancement we typically see
$T_{1}= $ 1 ns and $T_{2}=$ 0.6 ns \cite{Lee15} which would lead to
a variation in $I_{RRS}$ as shown in Figure \ref{Fig2}c in blue. The
maximum fraction $I_{RRS}/I_{total}$ is 0.3, which can only be
achieved well below saturation. For a cavity enhanced source the
fraction $I_{RRS}/I_{total}$ is close to unity (Figure \ref{Fig2}c,
red). We observe in Figure \ref{Fig2}b that at $\Omega =$ 0.83 GHz,
when $I_{total}$ is 0.65 of its maximum value, the RRS dominates the
spectrum.

\begin{figure}[h]
\includegraphics[width=100mm]{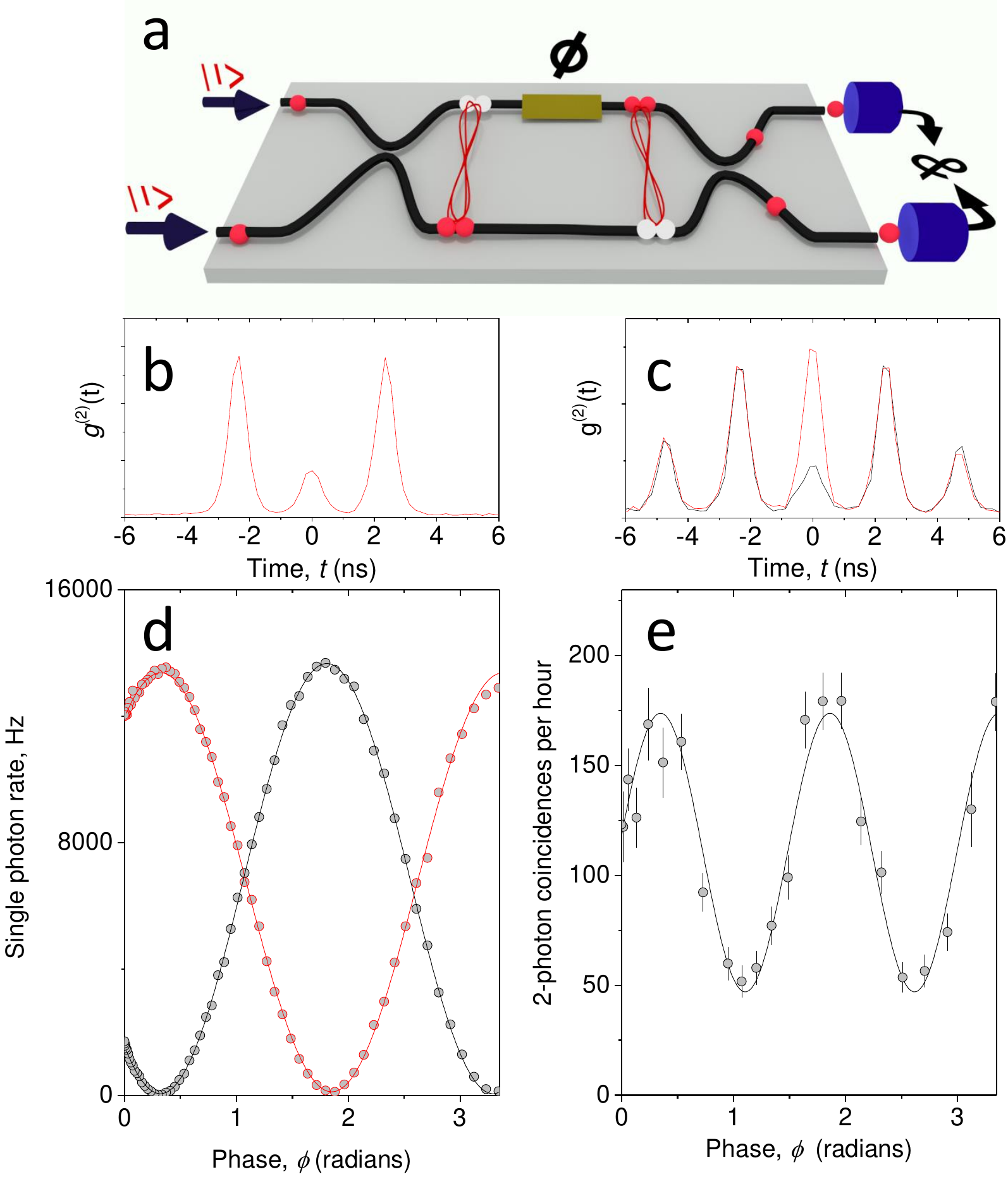}
\caption{\label{Fig3} \textbf{Deterministic excitation to create
on-demand indistinguishable photons and N00N states} (a) Schematic
of the photonic chip used to generate a 2-photon N00N state from two
single photons. (b) Auto-correlation measurement under pulsed
excitation with two laser pulses separated by 2.36ns. (c) Two-photon
interference between consecutive single photons with parallel
(black) and orthogonal polarisation (red). (d) The variation in
single photon detection rate at the output of the photonic chip as a
function of phase, $\phi$. (e) The two-photon coincidence detection
rate at the output of the photonic chip as a function of phase,
$\phi$.}
\end{figure}

Further confirmation of the character of the emitted light can be
obtained by two-photon interference measurements. Figure \ref{Fig2}d
shows the result of a continuous-wave two-photon interference
measurement \cite{Patel10, Ates09} at $\Omega = $ 0.83 GHz. Light
from source is passed to a fibre-optic Mach-Zehnder interferometer
with a delay of 10.4 ns and a half-wave plate (HWP) in one arm
(insert to Figure \ref{Fig2}d). This enabled photons emitted at a
time separation of 10.4 ns to take the two paths through the
interferometer and meet at the final coupler, $C_{B}$. Dependent on
the HWP photons meeting at this coupler can have parallel or
orthogonal polarization and an auto-correlation on the outputs of
the interferometer measures the degree of indistinguishability. The
difference in the two measurements is quantified by the interference
visibility $V_{\textrm{HOM}}(\tau) =
\left(g^{(2)}_{\perp}(\tau)-g^{(2)}_{\parallel}(\tau)\right)/g^{(2)}_{\perp}(\tau)$
which is shown in Figure \ref{Fig2}e. The maximum visibility
observed is 0.89, and the shape of the visibility plot is determined
by the first order coherence of the photons \cite{Cohen98}. We fit
this data \cite{Proux15, Cohen98} for a range of values of
$T_{2}/2T_{1}$. The calculation provides a remarkably good fit to
the data for $T_{2}/2T_{1}$ = 1.0. This shows the source is
delivering highly indistinguishable photons, and that the Purcell
effect has enhanced the RRS part of the spectrum.

Next, we discuss the operation of the source under pulsed optical
excitation to create on-demand single photons, indistinguishable
photon pairs and N00N states (Figure \ref{Fig3}). Optical pulses of
length 57 ps resonantly excite the transition and Rabi oscillations
in the detected RE are observed as a function of the incident field
amplitude. The source was driven with 0.71$\pi$- pulses to provide a
near-deterministic excitation and an optimal signal-to-background
level \cite{He13, Lee15}. When two pulses are used to excite the
source at a time separation of 2.36 ns the emitted photons can again
be interfered to determine their indistinguishability. For a pulsed
demonstration of 2-photon interference a more useful parameter is
$g$ \cite{Santori02} which is the probability of generating two
photons in either of the two pulses, divided by the probability of
generating two single photons. A measurement of $g$ = 0.167 $\pm$
0.005 is shown in Figure \ref{Fig3}b. For reference $g^{(2)}(0)=$
0.165 $\pm$ 0.004, where the probability of multi-photon emission
during a single laser pulse results in a deviation from the ideal
value. When these two photons interfere with with parallel or
orthogonal polarisation (black and red lines in Figure \ref{Fig3}c)
the difference in coincidences is indicative of a high degree of
indistinguishability. Following analysis \cite{Santori02}, we
determine a two-photon overlap
$|\langle\Psi_{1}|\Psi_{2}\rangle|^{2} = 0.90$.

Finally, we use this source to build a 2-photon N00N state, which is
the proto-typical quantum state used for phase-enhanced
measurement\cite{Dowling08, Boto00, Matthews09}. The two photons are
fed into a silicon oxy-nitride photonic chip as shown in Figure
\ref{Fig3}a which delivers sub-wavelength control of path lengths
and stability. Two indistinguishable photons interfere at the first
coupler and exit together in the superposition state
$|2\rangle_{A}|0\rangle_{B} + |0\rangle_{A}|2\rangle_{B}$. These two
paths then experience a relative phase shift, $\phi$, transforming
the state to $|2\rangle_{A}|0\rangle_{B} +
e^{2i\phi}|0\rangle_{A}|2\rangle_{B}$. Thus, super-resolution is
achieved because the phase shift introduced to the two-photon state
is twice that of a single photon, and can be measured by recombining
both paths on a final coupler.

Figure \ref{Fig3}d shows the single photon count rate at each output
of the photonic chip as a function of the phase shift $\phi$ when
light is only inserted to one of the inputs. The oscillations have a
visibility of 0.98, reflecting the balanced splitting of the
couplers and the two paths through the interferometer. When single
photons are fed into both inputs a measurement that post-selects
two-photon coincidences between the two outputs, Figure \ref{Fig3}e,
clearly displays a doubled rate of oscillation with phase, relative
to the single photon case. This is the signature of
super-resolution. The minimum coincidence rate observed in Figure
\ref{Fig3}e is determined by the value $g$. Losses in the optics and
detectors preclude the observation of phase super-sensitivity, so
this effect is limited to the post-selected measurements. However,
we anticipate that the future integration of single photon sources
directly onto the chip \cite{Murray15} could increase the efficiency
of the photon coupling.

In conclusion, we have demonstrated coherent excitation of a
cavity-QED enhanced emitter using the optimal axis for coupling in
and out light. We observe indistinguishable coherent photon
scattering even as the transition is driven near saturation. This
result shows how the combination of coherent excitation and
cavity-QED can lead to a bright and coherent photon source.

\parskip 5mm

$\textbf{Acknowledgements}$

\parskip 5mm

The EPSRC partly funded the MBE machine used to grow the sample. EM
and TM acknowledge support by the EU Marie Curie Initial Training
Network PICQUE, Grant No. 608062. JL acknowledges support from the
EPSRC CDT in Photonic Systems Development.

\bibliographystyle{apl}
\bibliography{bibliography3}

\end{document}